# A New Coordinated Beamformer for MIMO-based Ad Hoc Networks

Makan Zamanipour
Unit 5, No. 120, Jamshid Royan ST, PC. 1634776616
Tehran, Iran
makan.zamanipour.2015@ieee.org

Mohammadali Mohammadi
Shahrekord University
Shahrekord, Iran
m.a.mohammadi@eng.sku.ac.ir

*Abstract* — **The study proposes a new scheme for MIMO-based ad hoc networks. This is accomplished, while using the *Interference Driving Technique* (IDT) over Nakagami-*m* fading channels with perfect channel state information at both the transmitter and receiver. The use of this technique is proposed to decrease the impact of all the unwanted interferences, routinely caused by the overlap of the defined radio transmission ranges related to the used nodes. Indeed, IDT is utilized as a coordinated beamformer in a cooperative scheme, according to mean squared error criterion. The methodology and also analytical results are conducted to prove the aptitude of the paper.**

## I. INTRODUCTION

MIMO-based ad hoc networks have an undeniable role in wireless networks. One of the main problems in these networks is the overlap between the nodes and their defined radio transmission ranges. In other words, the more overlaps, the less quality of service, e.g. the packet throughput. To overcome this problem, a special *Interference Driving Technique* (IDT) according to [1] as the coordinated beamformer is proposed.

### A. Motivations

The main aim is to handle the dual unwanted interferences. This is realized, while establishing a number of coordinated beamformers according to IDT ([1]) at the transmitter of each node, associated with the number of overlaps. For instance, if we separately have any overlap for the node A related to the nodes B and C, two beamformers related to both of them should be separately and linearly established at the transmitter of the node A, according to our proposed scheme.

Here, the beamformer is attained according to mean squared error criterion. This is calculated, while focusing on the related equations in more detail. The design problem is fairly analyzed with perfect channel state information at both the transmitter and receiver.

Let us say the word at the beginning of that: For a given network and the given nodes which can be randomly located, apart from the network configuration, a coordinated beamformer is established for each node. The paper evaluates the merit of the scheme, characterizing some criteria such as the probability of packet error.

### B. Previous studies

To highlight the motivations, and so as to be able to actualize the contributions of the paper, related works should be given as follows in more detail.

In [1], IDT was adequately introduced to manage only the interferences that were caused by unlicensed users on authorized users in cognitive radio mobile networks. In other words in [1], only some of the forenamed interferences were considered over Rayleigh fading channels whose equations are the simplest.

In [2], the necessary stochastic criteria were widely examined over flat un-correlated Nakagami-*m* fading channels. Here, we completely use its proven equations.

In [3], an analysis was fulfilled to arrange the nodes so that the system performance could be significantly improved. In [4], a scheme was proposed to optimize the defined radio range. In [5], the link-layer throughput capacity was perfectly examined, while defining some solid criteria. Indeed, these studies provided some solid investigations, only about the configuration of the network. See [6-10] for details.

### C. Our Contributions

The contributions of the paper are classified below, deeply regarding the discussed motivations:

1) The usage of IDT ([1]) is proposed to to mitigate the interferences in MIMO-based ad hoc networks as a new proposition. Indeed, this realization which is given in terms of mathematical novelties, has not been realized until now. In fact, the use of this technique are evaluated in more detail, to overcome the problems related to the radio transmission ranges as a novelty.

2) Unlike [1], IDT is also performed to mitigate all the dual interferences, while adding new mathematical expressions. In other words, in [1], the mentioned technique was only introduced to cope with the interferences on the authorized user, caused by the unlicensed user. In fact, the interference on the unlicensed user, caused by the authorized user was not considered. This is based upon the fact of cooperative schemes such as cognitive radio networks. In other words, all defined users are considerably serviced here, to keep the quality of service significantly.

3) IDT is stablished over Nakagami-*m* fading channels that has not been actualized until now as a more valued examination, regarding more realestic and hard conditions.

In brief, apart from the configuration of the network, for the given nodes which can be randomly arranged, this paper geniunly proposes a new coordinated beamformer for each node as a novelty.

*D. Notations*

The notations that are mainly used throughout the paper, are given as follows. Scalars and matrices are respectively and widely declared by non-boldfaced lower-case characters and boldfaced upper-case ones. The used operations $\|.\|_F$, $E(.)$, and $(.)^H$ are properly the Frobenius norm, expected value and the Hermitian operations. The dimension of all MIMO channels is defined in terms of $M$. Hereafter, we use the italic font to indicate the proven or introduced topics.

*E. Remainder of the paper*

The remainder of the paper is well sectionalized as follows. At first, problem description for the proposed beamformer is perfectly discussed in the second section, while providing an overview on IDT according to [1]. The methodology in connection with the study to have more justifications, is fairly described. Numerical results and conclusions are given in the third and the fourth sections in more detail, respectively.

## II. PROBLEM DESCRIPTION

In this section, at first an overview on IDT is provided according to [1]. The main problem and the proposed scheme are adequately described. Meanwhile, the methodology by which the aforementioned scheme can be fairly justified, is discussed in more detail to more clarify.

*A. An overview on IDT*

Let us write a few about IDT as an overview in this part. In [1], this was introduced to overcome the unwanted interference on the authorized user, caused by the unlicensed user in a cognitive radio mobile network. Indeed, any beamformer was established at the transmitter of only the unlicensed user. This was capable of handling its interference, while having a cooperation to deal with the transmitted signal from only the unlicensed user.

*B. Proposed Scheme*

In this part, the proposed coordinated beamformer is discussed.

As can be observed from the Fig. 1, we have 3 nodes A, B and C which are linearly located, as a prime example. The shown circles are related to the defined radio transmission ranges, associated with the above-mentioned nodes. The main signals are shown as boldfaced arrows, whereas the interference signals are perfectly depicted as dotted arrows in the figure.

In practice, we have overlaps between the illustrated circles. Here, the key point is that each node as a MIMO transmitter has some coordinated beamformers according to IDT ([1]). This is related to the defined neighbor nodes which are linearly, randomly and fairly located as an example in this study. Of course, they should be multiplied. Here, as an example, the equations are obtained for the circles and the nodes A and C. It can be undoubtedly generalized to each node in multi-node schemes with more nodes associated with common overlap regions. Here, $A_C$ is defined as a given place related to the considered radio range of the node A. This is analyzed in the specified overlap region by the node C, on which the node A wants to transmit a signal. Similarly, $C_A$ is defined as a given place related to the identified radio range of the node C. But unfortunately, both places are located in the overlap region.

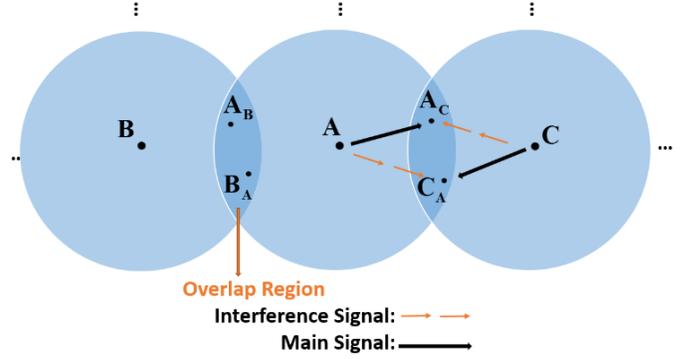

Fig. 1. Radio transmission range of the nodes A, B and C and their overlap regions.

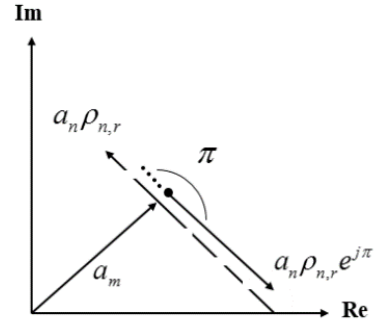

Fig.2. IDT: How to rotate the correlation coefficients (similar to [1]).

The received signals in $A_C$ and $C_A$ in the specified overlap region in the figure can be respectively expressed as:

$$\mathbf{Y}_{A_C} = \sum_I \frac{\mathbf{S}_{IA_C}\mathbf{M}_{I,J}\mathbf{X}_I}{\vartheta} + \mathbf{N}_{A_C}, \quad (1)$$
$$I = \{A,C\},\ J = \{A,C\},\ I \neq J,$$

and:

$$\mathbf{Y}_{c_A} = \sum_I \frac{\mathbf{S}_{IC_A}\mathbf{M}_{I,J}\mathbf{X}_I}{\vartheta} + \mathbf{N}_{c_A}, \quad (2)$$
$$I = \{A,C\},\ J = \{A,C\},\ I \neq J,$$

where $\mathbf{x}_I$ (I= {A, C}) is the transmitted signal at the transmitter of the *I*th node. $\mathbf{M}_{I,J}$ (I = {A, C}, J= {A, C} and J ≠ I) is the coordinated beamformer as IDT or the *Correlation Driver* ([1]) at the transmitter of the *I*th node, related to the defined overlap region by the *J*th node. $\mathbf{N}_{A_C}$ and $\mathbf{N}_{C_A}$ are respectively the noise matrices in $A_C$ and $C_A$. $\mathbf{S}_{IA_C}$ and $\mathbf{S}_{IC_A}$ are respectively the MIMO channel matrices between the *I*th node and $A_C$, and between the *I*th node and $C_A$. All the above-mentioned matrices are M-by-M. In addition, $\vartheta$ is the channel normalization coefficient that can be written as:

$$\vartheta = \sum_I \|\mathbf{M}_I\|_F^2, I = \{A,C\}. \quad (3)$$

Slightly similar to [1], the channel matrices of sizes $2M \times M$ are defined as:

$$\mathbf{S}_A = \begin{bmatrix} \mathbf{S}_{AC_A} \\ \mathbf{S}_{AA_C} \end{bmatrix}, \mathbf{S}_{A'} = \begin{bmatrix} \mathbf{S}_{CC_A} \\ \mathbf{S}_{CA_C} \end{bmatrix},$$
$$\mathbf{S}_C = \begin{bmatrix} \mathbf{S}_{CC_A} \\ \mathbf{S}_{CA_C} \end{bmatrix}, \mathbf{S}_{C'} = \begin{bmatrix} \mathbf{S}_{AC_A} \\ \mathbf{S}_{AA_C} \end{bmatrix}. \quad (4)$$

IDT in our proposed scheme is described here. How to use IDT as the *Correlation Driver* or *Rotator* ([1]) is given in Fig. 2, similar to [1]. $\alpha_m$ and $\alpha_n$ are the received signals associated with the main and the unwanted signals, respectively. Extra clarification is given as follows. According to mean squared error criterion and slightly similar to [1-eq. 20], the *Interference Driver Matrices* $\mathbf{M}_{A,C}$ and $\mathbf{M}_{C,A}$ at the transmitter of A and C can be written as (5) and (6), respectively. In fact, as completely given in [1], these matrices should be established to drive, and so as to rotate the correlation coefficients of the unwanted signals as their influences:

$$\mathbf{M}_{A,C} = \frac{\mathbf{S}_A^H}{(\mathbf{S}_A \mathbf{S}_A^H)} (\mathbf{\Theta}_A - \mathbf{M}_{C,A} \mathbf{S}_{A'}), \quad (5)$$

and:

$$\mathbf{M}_{C,A} = \frac{\mathbf{S}_C^H}{(\mathbf{S}_C \mathbf{S}_C^H)} (\mathbf{\Theta}_C - \mathbf{M}_{A,C} \mathbf{S}_{C'}). \quad (6)$$

(5) and (6) can be conveniently solved. Slightly similar to [1], $\mathbf{\Theta}_I$ (I= {A, C}) is the *Correlation Phase Rotator* matrix of size $2M \times M$ whose the *n*th and the *r*th element is $\rho_{n,r} e^{j\pi}$ which need to be fairly preserved, associated with $\mathbf{S}_I$ from (4). This is in connection with the cross-correlation matrix whose elements should be phase-corrected, according to [1].

Here, it should be noticed that $\pi$ is an example that can be generalized. In addition, it is only necessary to compute the expected value of $\rho_{n,r}$ in relevance to the cross-correlation matrices of fading channels. See [1] for extra investigation. Similar to [2-eq. A2], the above-mentioned elements should be obtained for Nakagami-m fading channels as:

$$E_{\mathbf{S}_D}\{\mathbf{S}_D \mathbf{S}_D^H\} = M \begin{bmatrix} \omega+\psi & \psi & \psi \\ \psi & .. & \psi \\ \psi & \psi & \omega+\psi \end{bmatrix}, D = \{A, A', C, C'\}, \quad (7)$$

where $\omega$ and $\sqrt{\psi}$ are respectively the variance and the mean of the elements of the defined channel matrices. $\mathbf{S}_D$ indicates the MIMO channels from (4) and $M$ is the dimension of MIMO channel matrices.

*C. Methodology*

In this part, the problem is concentrated in more detail as the methodology, to obtain the asserted beamformer.

Obviously, $\mathbf{M}_{A,C}$ is utilized to cope with the unwanted signal from the node A on $C_A$. Meanwhile, it should be noted that this has no effect on the channel norm related to $\mathbf{S}_{CC_A}$ and $\mathbf{S}_{CA_C}$. On the other hand, the channel norm associated with $\mathbf{S}_{AA_C}$ and $\mathbf{S}_{AC_A}$ is not affected by $\mathbf{M}_{A,C}$ and $\mathbf{M}_{C,A}$.

In contrast, by establishing both $\mathbf{M}_{A,C}$ and $\mathbf{M}_{C,A}$, all the dual unwanted and un-desired signals between A and C can be conveniently controlled. All the above-mentioned equations are written, while considering $\vartheta$ as the power normalization.

On the other hand, another beamformer as $\mathbf{M}_{A,B}$ should be designed about the node A related to the considered overlap region of the node B. As completely described before, all the interferences should be handled to keep the quality of service.

Now, to deal with the above-mentioned interferences completely, $\mathbf{M}_{A,C}$ and $\mathbf{M}_{A,B}$ should be mathematically, physically and linearly multiplied at the transmitter. To this end, we need to address the linear coordinated beamformer $\mathbf{M}_A$ as the equation (8), regarding *I* as the number of the adjacent nodes which have a linear effect:

$$\mathbf{M}_A = \prod_I \mathbf{M}_{A,I}, I = \{B, C, ...\}. \quad (8)$$

III. ANALYTICAL RESULTS

*MATLAB* and *Network Simulator-3* are used in this paper, as the most effective tools to perform the tests.

Some of the considered items in the simulations are mainly listed in this part. BPSK modulation is used in the simulations, while *M* is precisely set to 2. Of course, they are mainly changed in some figures to examine the system performance better. Meanwhile, $M$ can be fairly used in terms of both the diversity gain and multiplexing gain. Completely similar to [11], the length of each packet is set to 2304. Rayleigh fading channels are perfectly and basically produced as a special case of Nakagami-$m$ fading channels, such as [2]. The channels are un-correlated flat fading which are basically generated as slow fading channels, such as [2].

$C_{System}$ as the *System Capacity* generally and principally can be expressed as:

$$C_{System} = \log_2\{1+SNR\}, \quad (9)$$

where the signal to noise ratio (SNR) is in terms of the total transmit-power which shall be multiplied by the channel norm and the inverse of the noise variance, as completely described e.g. in [12].

Varying the number of the nodes, the system capacity is fairly depicted versus SNR in Fig. 3. As can be conveniently observed, the difference between the curves based upon this topic is notably clear. In this test, while increasing by a factor of 2 in the number of the nodes, the system capacity is degraded as 0.13 dB for SNR 5 dB as an example. It is clear that the less crowded network, the higher quality of service. In fact, this can be defined as a trade-off which depends on the aim of the design problem. In other words, for a given number of the nodes, the quality of service should be kept.

Now, with respect to [13, eqs. 6-8], the total packet error rate (PER) can be approximately disclosed as (10):

$$P_{packet} \approx 1 - \prod_{z=1}^{N_s}\left\{1-(1-p_e(z))^{\frac{N_z R_z}{K_z}}\right\}, \quad (10)$$

where:

$$P_e(z) \approx 1-(SER_z)\{d_{\min}(z)\}^{-1}. \quad (11)$$

According to [13], in (10-11), $d_{\min}(z)$, $N_z$, $R_z$, $K_z$, and $SER_z$ are respectively the minimum Euclidean distance, the number of bits, the code rate, the length and the symbol error rate related to the $z$th stream. Meanwhile, $N_s$ is the number of the used streams. The considered streams depend on how to use MIMO technique which can be used in terms of either multiplexing topic or diversity one.

Changing the distance between the used nodes, in Fig. 4 PER versus SNR is perfectly illustrated. The high distances in terms of meter can improve the above-mentioned error to more guarantee the quality of service. In other words, the probability of collision between the nodes can be dramatically decreased, and therefore the quality of service is well affected.

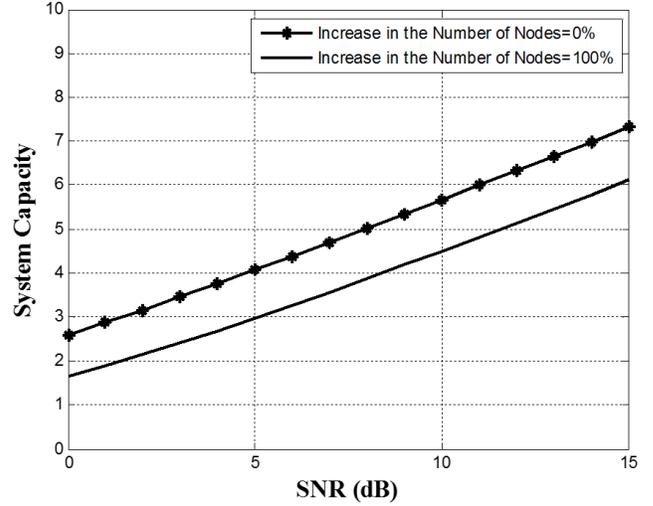

Fig.3. System capacity versus SNR, while making increase in the number of the nodes.

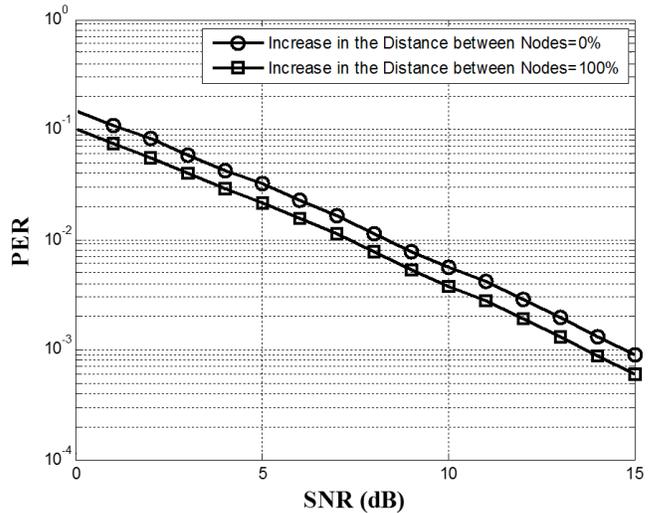

Fig.4. PER versus SNR, while changing the distance between the nodes.

A comparison between various modulations is made in Fig. 5, while showing PER against SNR. How the performance to be effected, is principally clear. In principle, the type of the used modulation has a key role in the system. Obviously, the system performance can be affected by Quadratic pulse shift keying modulation, better than binary pulse shift keying one. These above-mentioned modulations are illustrated in the figure as QPSK and BPSK, respectively.

While changing $M$, the bit error rate is fairly illustrated versus SNR regime in Fig. 6. Certainly, the increase in this parameter which is used in terms of the diversity gain, makes the probability of error to be basically and significantly improved. Indeed, the use of $M$ in terms of the diversity gain changes the slope of the curves.

It should be noted that the tests are performed under the assumption that we have ideally no interferences. In other words, the signal-to-interference-noise-ratio in this study is not analyzed.

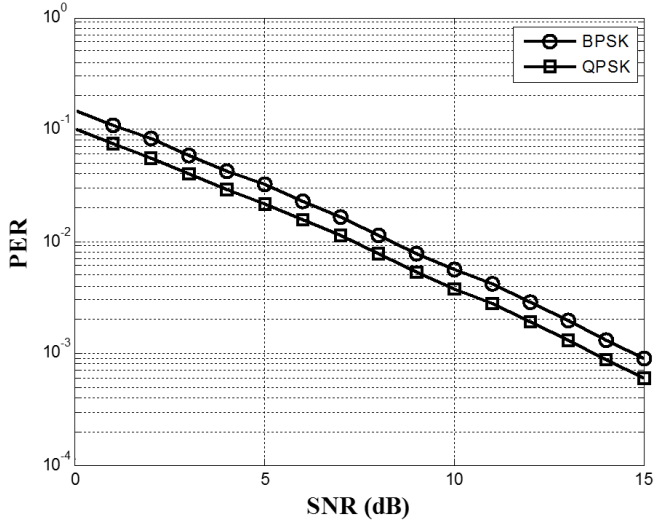

Fig.5. PER versus SNR, while changing the used modulation.

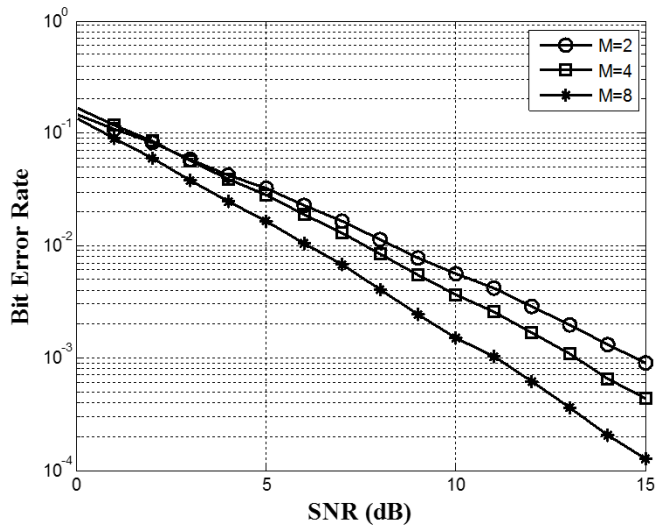

Fig.6. Bit error rate versus SNR, while changing the size of the MIMO channels.

In addition, the coordinated beamformer was proposed, while indirectly regarding the complexity of the system in terms of the increase in the number of the used nodes; otherwise solid references in which the forenamed topic was directly examined such as [14].

## IV. CONCLUSION

A new scheme for MIMO-based ad hoc networks was proposed. To this end, at first an overview on IDT was provided. Then, IDT as a coordinated beamformer was used to mitigate the impact of the unwanted interferences, produced by the overlap of the defined radio transmission ranges in the specified network. To more justify this, the used methodology was described in more detail. Also analytical results such as a closed-form expression for the packet error probability were essentially investigated.